*Effect of calcium stoichiometry on the dielectric response of $CaCu_3Ti_4O_{12}$ Ceramics*

**P. Thomas,[a,b] K. Dwarakanath,[a] K.B.R. Varma [b]***

[a] Dielectric Materials Division, Central Power Research Institute, Bangalore : 560 080, India.

[b] Materials Research Centre, Indian Institute of Science, Bangalore: 560012, India.

Abstract

$Ca_xCu_3Ti_4O_{12}$ (x = 0.90, 0.97, 1.0, 1.1 and 1.15) polycrystalline powders with variation in calcium content were prepared via the oxalate precursor route. The structural, morphological and dielectric properties of the ceramics fabricated using these powders were studied using X-ray diffraction, Scanning Electron Microscope along with Energy Dispersive X-ray Analysis, Transmission Electron Microscopy, Electron Spin Resonance (ESR) spectroscopy and Impedance analyzer. The X-ray diffraction patterns obtained for the x= 0.97, 1.0 and 1.1 powdered ceramics could be indexed to a body–centered cubic perovskite related structure associated with the space group Im3. The ESR studies confirmed the absence of oxygen vacancies in the ceramics that were prepared using the oxalate precursor route. The dielectric properties of these suggest that the calcium deficient sample (x= 0.97) has a reduced dielectric loss while retaining the high dielectric constant which is of significant industrial relevance.

*1. Introduction*

High-permittivity dielectric materials have attracted ever-increasing attention particularly for their practical applications in microelectronics such as capacitors and memory devices. The $CaCu_3Ti_4O_{12}$ (CCTO) ceramic belongs to a family of the type, $ACu_3Ti_4O_{12}$ (where A= Ca or Cd) and has been reported in the year 1967 [1]. This composition has been extended [2] and the family of titanates with the body centered cubic (bcc) structure has been reported having the general formula, $[AC_3] (B_4) O_{12}$, [where A= Ca,Cd,Sr,Na or Th; B = Ti or (Ti + $M^{5+}$), in which M= Ta, Sb or Nb; and C= $Cu^{2+}$or $Mn^{3+}$]. CCTO has attracted considerable attention recently due to its unusually high dielectric constant ($\varepsilon \sim 10^{4-5}$) which is nearly independent of frequency (upto 10 MHz) and low thermal coefficient of permittivity (TCK) over a wide range of temperatures (100-600K) [3,4]. Several explanations for the origin of high dielectric constant for CCTO have been proposed mainly based on bulk property contributions as against the microstructural features which in turn are affected by the ceramic processing conditions including sintering temperatures as well as the ambience [3,5-7]. Accordingly, the incidence of giant dielectric constants has been attributed to more than a single mechanism : (i) the barrier layer capacitance arising at twin boundaries [3]; (ii) disparity in electrical properties between grain interiors and grain boundaries [8-10]; (iii) space charge at the interfaces between the sample and the electrode contacts [11,12]; (iv) polarizability contributions from the lattice distortions [13]; (v) differences in electrical properties due to internal domains [14]; (vi) dipolar contributions from oxygen vacancies [15-16]; (vii) the role of Cu off-stoichiometry in modifying the polarization mechanisms [17] ; (viii) cation disorder induced planar defects and associated inhomogeneity [18] or (ix) nanoscale disorder of Ca/Cu substitution giving rise to electronic contribution from the degenerate $e_g$ states of Cu occupying the Ca site contributing to the high dielectric constant [19].

* Corresponding author : Tel. +91-80-2293-2914; Fax: +91-80-2360-0683.
E-mail : kbrvarma@mrc.iisc.ernet.in (K.B.R.Varma)

Though various mechanisms for the origin of giant dielectric response have been proposed, the reports based on the twin boundaries [3], grain boundaries [8], electrode-sample interfaces [11,12], and the first principle calculations [20,21] provided the strong evidence of extrinsic effects associated with CCTO. Moreover the internal barrier layer capacitor (IBLC) model representing semiconducting grains and insulating grain boundaries [5] confirmed the electrical heterogeneities in the microstructure of CCTO, and has been widely accepted as the most likely explanation for the abnormal dielectric response in CCTO. These inhomogeneous features at the grain boundaries were also supported by current–voltage measurements [22].

To further elucidate the mechanism and also to improve the dielectric properties, cation doping effect on CCTO has been studied. Each dopant had its own preferential substitutional site when incorporated into the CCTO lattice. Certain dopants have critical influence on electrical barriers at the grain boundaries and a reduction in the number of charge carriers in the bulk grains as well [23]. The doping studies in CCTO were mainly attempted to improve its dielectric loss characteristics by controlling the chemistry and structure of interfacial regions at the grain boundaries [24-29]. Apart from the doping effect on the specific cation site, Ti deficient [30], Ti excess [31], Cu excess [32], and the Cu deficient [33] effects have been investigated on the dielectric response of CCTO. The Ti-deficient in CCTO has exhibited lower dielectric constant [30], while the Ti-rich CCTO ceramic [31] exhibited higher dielectric response. The Cu ions in CCTO appear to play a very important role as the segregation of copper oxide at the grain boundaries is believed to be responsible for the high resistance associated with the grain boundary [32-38]. However, the effect of stoichiometry, i.e., the relative amount of Ca, Cu and Ti is not well studied. As the Ca and Cu occupy the same A site in CCTO, the Ca stoichiometry studies are very important and need to be understood in the light of its dielectric properties.

This article reports the details pertaining to the effect of Ca variation on the dielectric properties of CCTO ceramics fabricated employing the powders obtained by the oxalate precursor route.

## 2. Experimental

The precursors $Ca_xCu_3(TiO)_4(C_2O_4)_5(CO_3)_3 \bullet CO_2$, (x= 0.90, 0.97, 1.0, 1.1 and 1.15) with varied *x* were prepared by employing the oxalate precursor route that is reported elsewhere [39]. Initially, the titania gel was prepared from the aqueous $TiOCl_2$(0.05M) by adding $NH_4OH$ (aq) (at 25°C) till the pH reaches ~ 8.0 and washed off $NH_4Cl$ on the filter funnel. To this titania gel [0.4 moles $TiO_2xH_2O$ (where 92<x <118)], powdered $H_2C_2O_4 \bullet 2H_2O$ was added and slightly heated to 40°C with continuous mixing without the addition of water. The amount of water present in the titania gel was sufficient for dissolving the whole of the solid present and a clear solution was obtained. To this clear solution, calcium carbonate (where calcium molar ratio is x = 0.90, 0.97, 1.0, 1.1 and 1.15) was added and stirred. The solution remained clear without any precipitate formation. This solution was cooled to 10°C to which cupric chloride dissolved in acetone+water (80/20) was added and stirred continuously. The thick precipitate was separated out by further addition of acetone. The precipitate was filtered, washed several times with acetone to make it chloride-free and dried in air. The precursors thus prepared were isothermally heated above 680°C to get the ceramic powders of $Ca_xCu_3Ti_4O_{12}$ (x = 0.90, 0.97, 1.0, 1.1 and 1.15), which are designated as $Ca_{0.90}$, $Ca_{0.97}$, $Ca_{1.0}$, $Ca_{1.1}$ and $Ca_{1.15}$. These notations have been used throughout the text henceforth.

The X-ray powder diffraction studies were carried with an X'pert diffractometer (Philips, Netherlands) using Cu K$\alpha_1$ radiation ($\lambda$ = 0.154056 nm) in a wide range of 2θ (5° ≤ 2θ ≤ 85°) with 0.0170 step size using the 'Xcelerator' check program. The phase pure CCTO powders were cold-pressed into pellets of 12mm in diameter and 3mm in thickness using 3% poly vinyl alcohol (PVA) and 1% polyethylene glycol as the binders. The green pellets were then sintered at 1130°C/4h. The grain size measurements were carried out using the Sigmascanpro image analysis software. The thin foil specimens for TEM were prepared from polycrystalline bulk samples by mechanical thinning and finally by Ar-ion milling using the precision polishing system (PIPS,GATTAN,691,CA) with liquid $N_2$ cooling to minimize the ion beam damage. Electron Diffraction and Transmission Electron Microscopic (TEM) studies were carried out on the sintered ceramics using FEI-Technai TEM (G-F30, Hillsboro, USA). Scanning electron microscope (FEI-Technai TEM-Sirion) equipped with Energy-Dispersive X-ray spectroscopy (EDX) capability was used to observe the microstructure and the composition of the sintered pellets. The densities of the sintered pellets were measured by the Archimedes principle using xylene as the liquid medium. The capacitance measurements on the electroded (sputtered gold and silver) pellets were carried out as a function of frequency (100Hz–1MHz) using an impedance gain-phase analyzer (HP4194A).

## 3. Results and Discussion

### 3.1. X-Ray diffraction studies

The X-ray diffraction patterns obtained for all the compositions of $Ca_xCu_3Ti_4O_{12}$ (x = 0.90, 0.97, 1.0, 1.1 and 1.15) are depicted in the fig.1(a-e). The diffraction patterns obtained for $Ca_{0.97}$, $Ca_{1.0}$, and $Ca_{1.1}$ samples as shown in the fig.1 (b-d) could be indexed to a body–centered cubic perovskite related structure of space group Im3. Whereas, the diffraction patterns obtained for the $Ca_{0.90}$ and $Ca_{1.15}$ samples (fig.1 a&e) did not show monophasic nature of CCTO. The $Ca_{1.15}$ sample (fig.1a) shows the presence of secondary phases pertaining to $CaTiO_3$, $TiO_2$ and CuO, whereas the $Ca_{0.90}$ sample (fig.1e), revealed the presence of only $TiO_2$ and CuO.

### 3.2 Microstructural studies (SEM/TEM)

The CCTO compositions ($Ca_{0.97}$ $Ca_{1.0}$ and $Ca_{1.1}$) which were monophasic were chosen and sintered at $1130^oC/4h$ for further investigations. The SEM micrographs were recorded for these pellets after thermal etching. Though all the samples are of monophasic in nature as revealed by XRD analyses, there is a significant difference in the microstructure. The microstructure of the samples $Ca_{1.0}$ (fig.2a) exhibited fascinating microstructure in which each grain is surrounded by exfoliated sheets of Cu-rich phase. The fig.2b shows the SEM picture at high magnification at the grain boundary region exhibiting such exfoliated sheets of Cu-rich phase. The EDX studies carried out on the grain and the grain boundary regions for the $Ca_{1.0}$ sample (fig.2 c and d) clearly indicate that the grain boundary region is rich in Cu. Hence, the microstructural features are broadly described as the stoichiometric grains embedded in a "pool" of Cu-rich phase. The microstructure that is evolved for the $Ca_{0.97}$ sample (fig.3a) more or less resembles that of the $Ca_{1.0}$ (fig.2a), but the segregation of such exfoliated sheets of cu-rich phase is less and thinner than that of $Ca_{1.0}$ sample (fig.2a). The EDX data for the composition of the grain and the grain boundary regions are shown in the fig.3 (b and c) for the $Ca_{0.97}$ sample. Fig.4(a-c) shows the SEM photographs of the pellets along with the EDAX analysis for the $Ca_{1.1}$ sample. The micrograph indicates that there is no discernible grain boundary phases. Since the microstructure reported is composition dependent, some of the samples may be poor in stoichiometry. We had made attempts to sinter the green pellets at lower temperatures ($<1130^oC$). Since the ceramics sintered at lower temperatures did not exhibit reasonable density, keeping the dielectric properties in view, these samples were sintered at high temperatures to obtain reasonably high density and consequently high dielectric constant. The reason for the kind of microstructure that is encountered in the present studies is attributed to high sintering temperature associated with the present samples. The EDX analyses carried out on the grain (fig.4 b) and for the overall region (fig.4c) for the $Ca_{1.1}$ sample suggest the composition to be almost the same. The table

1 illustrates the results on compositional analyses carried out by EDX for all the samples at different locations: grain, grain boundary and the overall regions.

The grain size measurements were carried out by line intercept method using the image analysis software. The grain size distribution obtained is given in the fig. 5(a-c). For $Ca_{1.0}$ sample (fig.5a), the grain sizes vary from 10-100 μm with most grains lying in the 30-70 μm range. The population of grains in the range of 20-30 μm and 70-100 μm are significantly low (around 10% each). In the case of $Ca_{0.97}$ sample, the population of grains in the range of 5-10μm and 30-40μm are also significantly low (around 10% each) with most grains lying in the 10-20μm range (51%). The noticeable difference is that, the grains in the 10-20μm range is totally absent for the $Ca_{1.0}$ sample. The $Ca_{1.1}$ samples (fig.5c), had grains in the range of 5-30μm and the population of grains is in the range of 10-20μm (57%) which is slightly higher than that of $Ca_{0.97}$ samples (51%). Overall, it is observed that, the $Ca_{0.97}$ sample (fig.5b), has small grains as compared to that of $Ca_{1.0}$ sample (fig.5a).

The TEM studies were carried out to probe domain features or local distortion, if any including stacking faults or 2D defects in CCTO as observed by the others [14,40,41]. The $Ca_{0.97}$ sample which exhibited smaller grains and thinner CuO segregation at the grain boundaries was chosen for these studies. The transmission electron microscopy (TEM) on the $Ar^+$ ion-thinned specimen ($Ca_{0.97}$) were carried out. Thin samples (<1000 $A^o$) suitable for the electron diffraction (ED) were prepared from the bulk polycrystalline $Ca_{0.97}$ ceramic. Fig.6a shows the SAED pattern with the zone axis as [111]. The ED pattern recorded along the [111] zone axes are consistent with the space group Im3. There are no elongations, streaks or distortion discernible for the diffraction spots indicating the absence of stacking faults or 2D defects. Fig.6b indicates the high resolution electron microscopy (HRTEM) image recorded for the $Ca_{0.97}$ sample along the [111] zone axis. The micrograph does not strikingly show the presence of any lattice defects such as stacking faults or dislocations. The micro domain features within the grains [14] and the lattice misfit [40] between the domains in a single grain causing local distortion were totally absent for the CCTO prepared via soft chemistry route. Zhu.et al [19] had observed that nanoscale disorder of Ca/Cu substitution giving rise to electronic contribution from the degenerate $e_g$ state of Cu occupying Ca site facilitate high dielectric constant. This needs to be substantiated by making detailed studies using quantitative electron diffraction and extended x-ray absorption.

### *3.3 Electron Spin Resonance (ESR) studies.*

Fig. 7 illustrates the ESR spectra recorded for the sintered ceramics ($Ca_{1.0}$, $Ca_{0.97}$ and $Ca_{1.1}$) at room temperature using an X-band (9.839) Bruker (Germany) spectrometer. It is well established that oxygen vacancies and /or intrinsic defects in oxides have been demonstrated to play a crucial role in the physical properties of different classes of transition metal oxides [41-44]. The broadening of the ESR line-width of CCTO when annealed in argon atmosphere is reported to arise from the oxygen vacancies [35,45]. It decreases on reannealing in oxygen atmosphere. The ESR spectra obtained for all the samples show single-line symmetric signal of g=2.15 having no hyperfine structures and nearly Lorentzian line shape. The appearance of the single-line spectra of g=2.15 for all these samples indicates the formation of copper (II) with square planar coordination. The spectra obtained for all the samples are quite comparable to those reported for CCTO [35,45,46], which have been assigned to copper (II) in square planar coordination. Hence, based on the ESR studies, the presence of oxygen vacancies in the ceramics that were fabricated using the oxalate precursor route is ruled out.

### *3.4 Dielectric studies*

The frequency dependent dielectric constant ($\varepsilon_r'$) and dielectric loss (D) at room temperature for the samples, $Ca_{0.97}$, $Ca_{1.0}$ and $Ca_{1.1}$ are shown in the fig.8(a and b). The dielectric constant value obtained for $Ca_{1.0}$ samples is around 25300 at 100Hz and decreased to 6747 as the frequency increased to 1MHz. The $Ca_{1.1}$ sample, which has higher calcium content (by 0.1 mol%) exhibited a dielectric constant value of around 13160 at 100Hz and 8557 at 1MHz. The $Ca_{1.1}$ sample shows low frequency dispersion as compared to that of $Ca_{1.0}$ samples. Interestingly, the $Ca_{0.97}$ sample, which is calcium deficient by 0.03 mol%, exhibited very high dielectric constant as compared to that of other samples ($Ca_{1.1}$ and $Ca_{1.0}$) at all the frequencies under study. The value of dielectric constant for $Ca_{0.97}$ sample vary from 32,900 to 27,980 in the 100Hz-1MHz range and the loss varies from 0.10 to 0.087. At 1kHz, the dielectric constant value obtained is 31470 and the loss is around 0.087. This sample also exhibited low frequency dispersion (100Hz-1kHz) as compared to that of $Ca_{1.0}$. The dielectric loss did not show any relaxation at low frequency, though there is a relaxation at high frequency. The dielectric loss (at 1kHz) values obtained for $Ca_{1.0}$ and $Ca_{1.1}$ are 0.14 and 0.10 respectively. It is to be noted that the calcium deficient sample ($Ca_{0.97}$), exhibited low dielectric loss (0.087 at 1kHz), while retaining the high dielectric constant. It is known in the literature that various dopants were tried [23-29] to bring down the dielectric loss in CCTO. It was also observed that, reduction in the loss in CCTO also affected the dielectric constant to a large extent. However, in this work, without any dopant, only by controlling the chemistry and engineering the interfacial regions at the grain boundaries, the dielectric loss was suppressed remarkably while retaining the giant dielectric constant.

The temperature dependent ($50^o$-$200^o$C) characteristics of dielectric constant ($\varepsilon_r'$) and dielectric loss (D) were monitored at three different frequencies (1kHz,10kHz,100kHz) for all the samples ($Ca_{0.97}$ $Ca_{1.0}$ and $Ca_{1.1}$) under investigation which are depicted in figs. 9 and 10 respectively. The dielectric constant decreases as the frequency increased from 1kHz to 100kHz in the $50^o$-$200^o$C temperature range. The dielectric constant increases steadily with increasing temperature for all the samples. The dielectric constant value at 1kHz has increased from 23400 to 62200 as the temperature increased from $50^{o}C$ to $200^{o}C$ for the $Ca_{1.0}$ sample and for the $Ca_{0.97}$ sample, the increase is from 29400 to 79000. In the case of $Ca_{1.1}$ sample, it has increased from 11700 to 20700 when the temperature is increased from $50^{o}C$ to $200^{o}C$. The temperature dependent characteristics of dielectric loss (D) at various frequencies for all the samples ($Ca_{0.97}$ $Ca_{1.0}$ and $Ca_{1.1}$) are shown in the fig 10. The frequency independent dielectric loss was observed upto $75^{o}C$ and subsequently it decreased with increasing frequency for all the samples under study. The dielectric loss increases rapidly beyond $100^{o}C$ for all the samples indicating the dominance of leakage current at elevated temperatures.

The higher dielectric constant exhibited by $Ca_{0.97}$ ceramic is rationalized by invoking essentially extrinsic effects. The $Ca_{0.97}$ sample does not exhibit broadening in the ESR line width. Hence, the large dielectric constant associated with this sample could not be attributed to the presence of oxygen vacancies. The transmission electron microscopy also did not reveal any lattice defects such as stacking faults or dislocations. The micro domain features and the lattice misfit were also totally absent in these ceramics. Hence, these findings have provided the strong evidence of extrinsic effects in CCTO. The SEM analysis very clearly indicated that there is a remarkable difference in the microstructural features for these ceramics and is evidenced that the dielectric response in these ceramics has strong microstructure dependence. The increase in the dielectric constant values with the grain size supports the argument for an internal boundary layer capacitor model to explain the dielectric behaviour of CCTO [11]. However, in this work, it is observed that, the $Ca_{0.97}$ sample which exhibited very high dielectric constants has smaller grains (fig.5b) than that of $Ca_{1.0}$ (fig.5a). Hence, there is no correlation that could be clearly drawn between the grain size and the dielectric constant based on these results. Similar observations were reported in the literature [47]. It is well know that the Cu ions in CCTO play a very important role as the segregation of copper oxide at the grain boundaries is believed to be responsible for the high resistance associated with the grain boundary [17,32-36].

It is to be noted that, Capsoni et al [35] have reported high dielectric constants for the samples containing small amounts of CuO segregation at the grain boundaries. This was further supported by Yeoh et al [36] wherein, it was observed that the higher amounts of CuO segregation at the grain boundaries exhibited low dielectric constants as compared to that of the samples with lower CuO segregation. Interestingly, the microstructure evolved (fig.3b) for the calcium deficient samples ($Ca_{0.97}$) exhibited less CuO segregation at the grain boundaries. This sample exhibited very high dielectric constant as compared to that of the $Ca_{1.0}$ sample which has higher amounts of CuO segregation (fig.2b) at the grain boundary. Though the sintered ceramics, subjected to XRD analysis (not reported in this manuscript) did not reveal the presence of any detectable secondary phases pertaining to CuO phases, the formation of negatively charged species as reported in the literature [37] existing at the grain boundaries due to the oxidation of the secondary phase (minor amount) during cooling after sintering cannot be ruled out. However, the secondary phase that was detected using EDX (atomic %) at the grain boundaries is close to that of CuO which is more stable than $Cu_2O$ in air at room temperature [38]. Also as revealed by SEM micrograph, the grain boundary layers (exfoliated sheets) seem to be less in thickness than that observed in $Ca_{1.0}$ which might be the reason for the higher dielectric constant associated with the $Ca_{0.97}$ sample. It has been reported in the literature that for CCTO, $R_{gb} \gg R_g$ and $C_{gb} \gg 10 C_g$ [8,10]. Therefore, the effective dielectric constant of the sample at frequencies much lower than the relaxation frequency is given by

$$\varepsilon'_{eff} = \frac{C_{gb}}{C_o} \qquad (1)$$

$\varepsilon'_{eff}$ is determined by the ratio between grain boundary capacitance $C_{gb}$ and empty cell capacitance $C_o$. If it is assumed that grain and grain boundary form two layer capacitor with a thickness ($d_g+d_{gb}$), where $d_g$ and $d_{gb}$ are the thickness of the grain and the boundary layer respectively, from eqn(1), one obtains $\varepsilon'_{eff} = \varepsilon_{gb}\left[\dfrac{d_g + d_{gb}}{d_{gb}}\right]$, here $\varepsilon_{gb}$ is the dielectric constant of the grain boundary layer. Therefore, even a small $\varepsilon_{gb}$ would lead to a giant dielectric constant $\varepsilon'_{eff}$ if the ratio $(d_g + d_{gb})/d_{gb}$ is large. Hence, it is clear that $\varepsilon'_{eff}$ is dependent upon the microstructure and could therefore be maximized by enhancing the grain growth (increasing $d_g$) or by thinning down the grain boundary, or by forming the grain boundary phase with high $\varepsilon'_r$. The above concept was adopted to rationalize the present dielectric data and has been observed that the ratio $(d_g + d_{gb})/d_{gb}$ obtained is higher for the $Ca_{0.97}$ sample than that of $Ca_{1.0}$ sample. Hence, it is concluded that, the samples with higher concentrations of CuO segregation exhibit low dielectric constant and those with low CuO segregation exhibit high dielectric constant which is in close agreement with that reported in the literature [35,36]. From these observations, it is concluded that the dielectric response observed in these samples has strong microstructure dependence especially, CuO segregation at the grain boundaries.

In order to have an insight into the relatively low dielectric loss observed for $Ca_{0.97}$ ceramic, the ac conductivity measurements were carried out in the frequency of interest for all the three samples under investigation. The frequency dependence of the ac conductivity ($\sigma_{ac}$ in log-log scale) at room temperature is shown in fig 11. The conductivity for all the three samples is found to increase with increasing frequency. However, in the case of $Ca_{0.97}$ sample, the rate of increase in $\sigma_{ac}$ is not that rapid and is lower than that of the other two samples especially at high frequencies. The mechanism of conduction in stoichiometric CCTO is attributed to the hopping of the charge carriers between the $Cu^{2+}$ and $Cu^{3+}$[17]. However, recently the conductivity mechanism proposed based on detailed experimental investigations point towards n-type electron hopping between $Ti^{3+}$ and $Ti^{4+}$ [48,49]. One of the main reasons for low loss associated with $Ca_{0.97}$ sample might be due to the less available oxygen vacancies vis-à-vis electron hopping between $Ti^{3+}$ and $Ti^{4+}$.

*4. Conclusions*

CCTO ceramics with different calcium contents were fabricated from the fine powders obtained by oxalate precursor route. The samples corresponding to the compositions $Ca_{0.97}$, $Ca_{1.0}$ and $Ca_{1.1}$ were monophasic in nature as confirmed by the X-ray diffraction studies. The calcium deficient sample ($Ca_{0.97}$) exhibited very high dielectric constant accompanied by low loss. This work clearly demonstrates that by controlling the chemistry and structure of interfacial regions especially at the grain boundaries, the loss could be suppressed significantly while retaining the giant dielectric constant. The ESR and TEM studies evidently suggest that the dielectric response in CCTO ceramics is of extrinsic origin. The calcium deficient samples ($Ca_{0.97}$) exhibiting low level of CuO segregation at the grain boundaries give rise to high dielectric constant, which is in line with that reported in the literature [31,32].

|  | EDAX analysis | | | | | | | | |
| --- | --- | --- | --- | --- | --- | --- | --- | --- | --- |
| Samples | Grain region | | | Grain boundary | | | Overall region | | |
|  | Ca | Cu | Ti | Ca | Cu | Ti | Ca | Cu | Ti |
| Ca $_{1.0}$ | 1.0 | 3.5 | 4.54 | 1.0 | 44.6 | 4.29 | 1.0 | 6.21 | 4.22 |
| Ca $_{0.97}$ | 1.0 | 3.41 | 4.27 | 1.0 | 19.03 | 4.28 | 1.0 | 4.01 | 4.38 |
| Ca $_{1.1}$ | 1.0 | 3.0 | 4.10 |  | ---- |  | 1.0 | 3.26 | 4.2 |

Table.1 The EDX analysis carried out on the grain, grain boundary and the overall regions for three different compositions.

*Figure captions:*

Figure. 1. X-ray diffraction patterns for $Ca_xCu_3Ti_4O_{12}$ where (a) x=0.9 ($Ca_{0.90}$), (b) x=0.97 ($Ca_{0.97}$) (c) x=1.0 ($Ca_{1.0}$) (d) x=1.1 ($Ca_{1.1}$) (e) x=1.15 ($Ca_{1.15}$) and (f) CCTO from the ICDD data file card no. 01-075-1149

Figure. 2. (a) Scanning electron micrographs of the $Ca_{1.0}$ sample sintered at $1130^oC/4h$, (b) the high magnification image showing the exfoliated sheets of Cu rich phase at the grain boundary region, EDX spectrum of (c) the grain and (d) the grain boundary region.

Figure. 3. Scanning electron micrographs of the $Ca_{0.97}$ sample (a) ceramic sintered at $1130^oC/4h$, (b) the high magnification image showing the Cu segregation at the grain boundary region, EDX spectrum of (c) the grain and (d) the grain boundary region

Figure. 4. (a) Scanning electron micrographs for the $Ca_{1.1}$ sample sintered at $1130^oC/4h$, EDX spectrum of (b) the grain and (c) the overall region

Figure.5. The grain size distribution obtained by line intercept method for (a) $Ca_{1.0}$, (b) $Ca_{0.97}$ and (c) $Ca_{1.1}$ samples sintered at $1130^oC/4h$.

Figure. 6 (a) [111] zone-axis selected area electron diffraction pattern of $Ca_{0.97}$ (b) HRTEM lattice image of $Ca_{0.97}$ sample along [111] zone axis

Figure. 7 ESR spectra recorded for different samples at room temperature

Figure 8. Frequency dependence of room-temperature (a) dielectric constant ($\varepsilon_r'$) and (b) dielectric loss (D) for the pellets sintered at 1130°C/4h.

Figure. 9 Temperature dependence of dielectric constant ($\varepsilon_r'$) measured at various frequencies for (a) $Ca_{1.0}$, (b) $Ca_{0.97}$ and (c) for $Ca_{1.0}$ samples

Figure. 10 Temperature dependence of dielectric loss (D) measured at various frequencies for (a) $Ca_{1.0}$, (b) $Ca_{0.97}$ and (c) for $Ca_{1.0}$ samples

Figure. 11 Frequency dependence of ac conductivity for $C_{1.0}$, $Ca_{0.97}$ and $Ca_{1.1}$ samples.

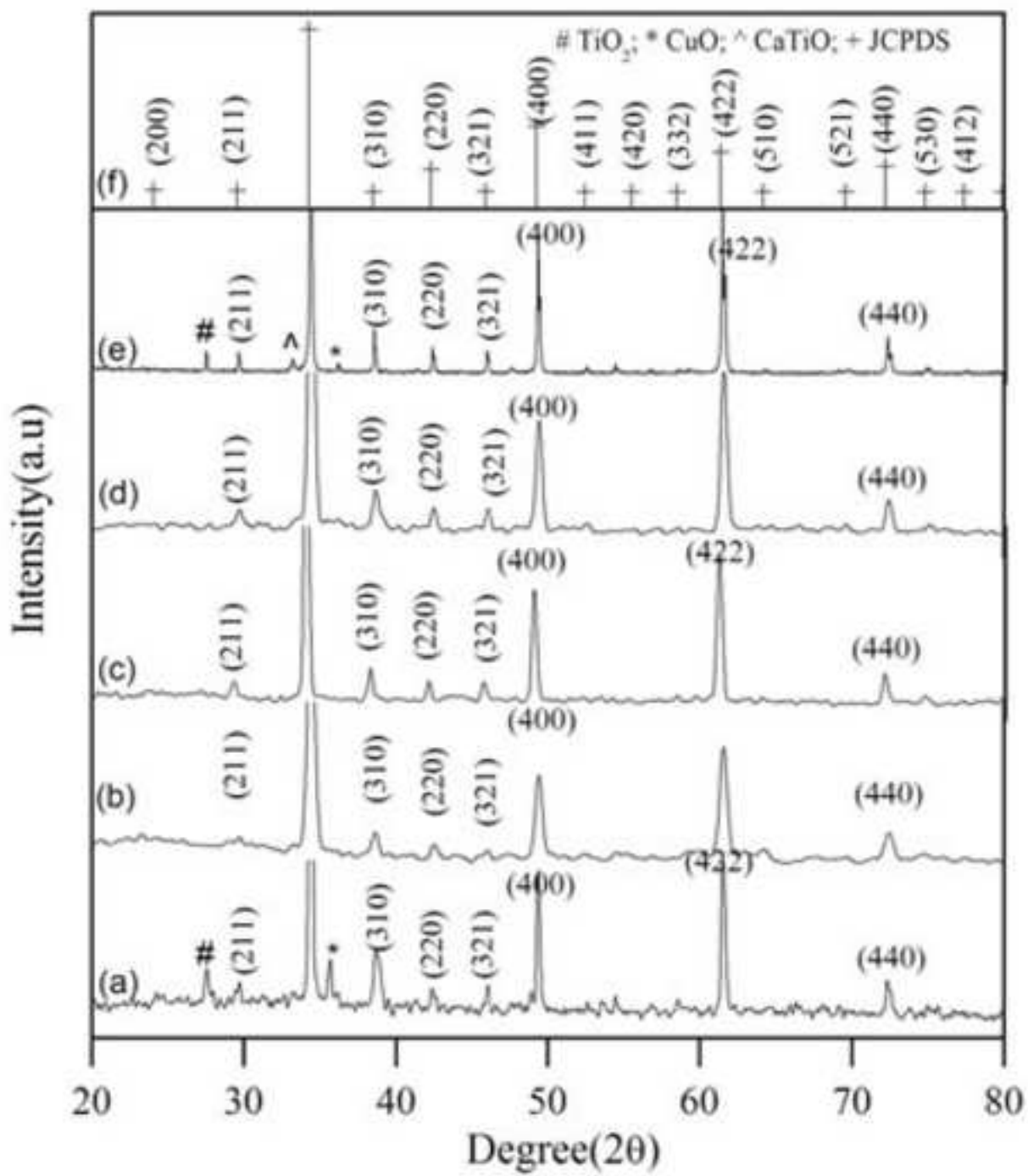

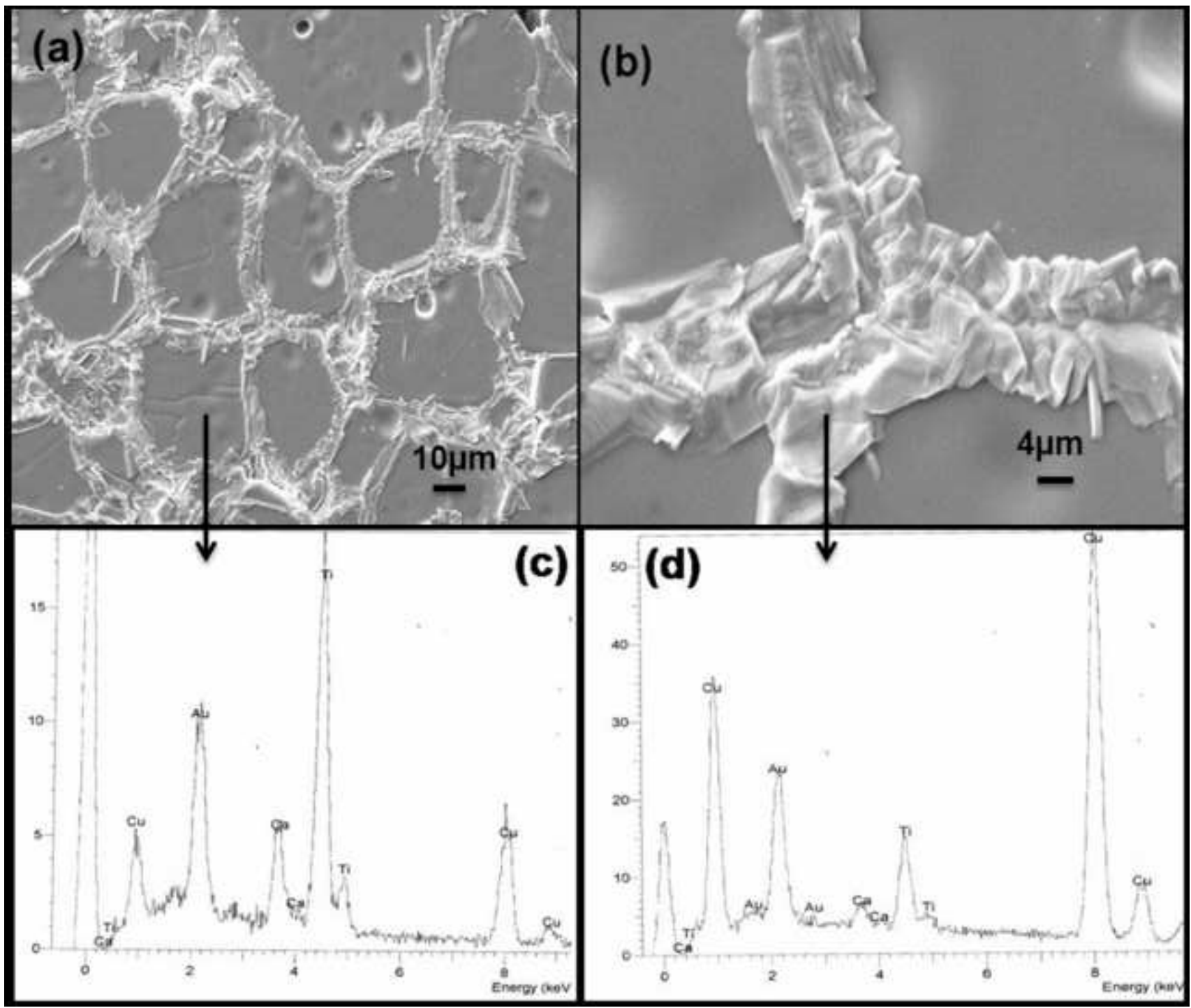

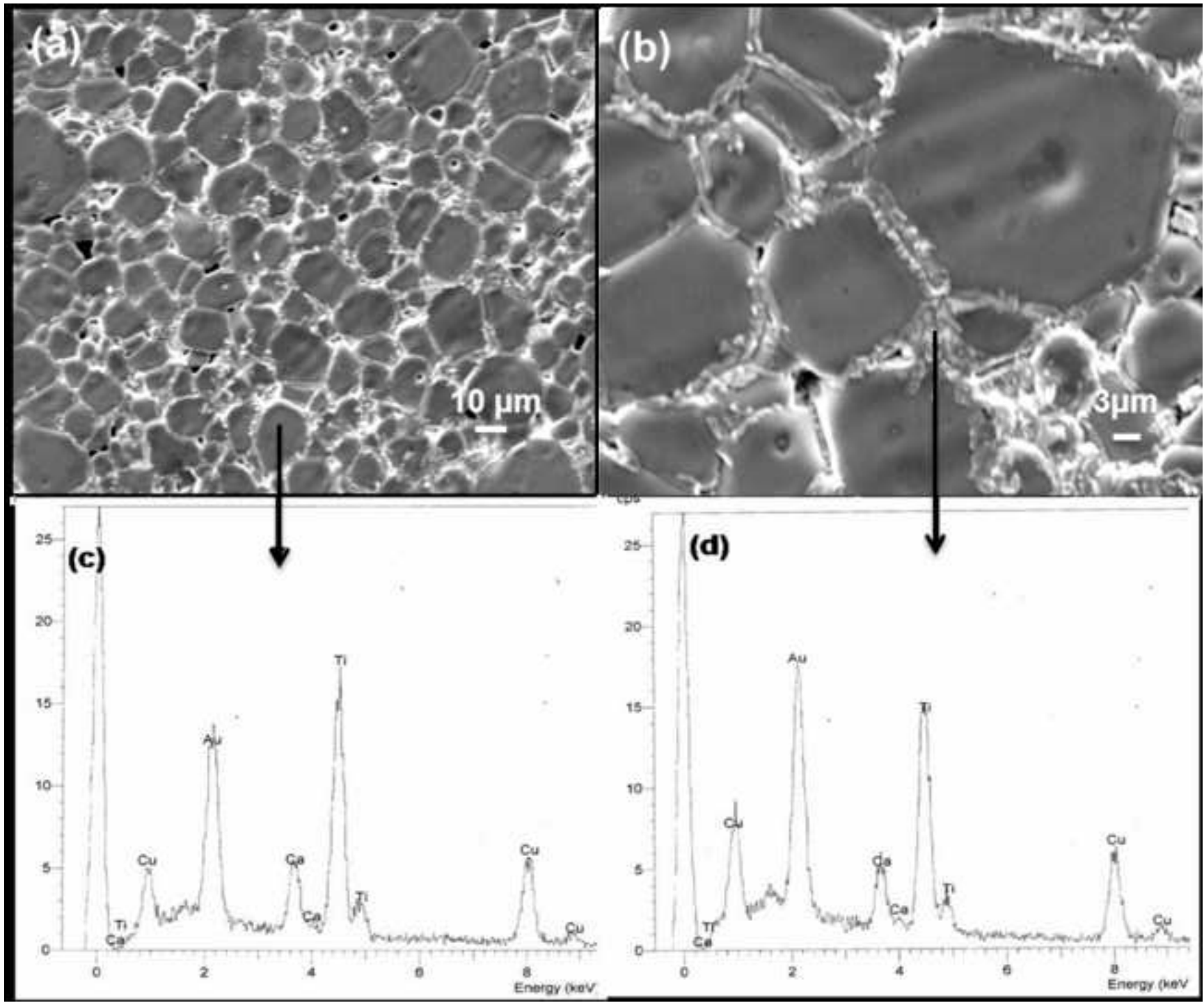

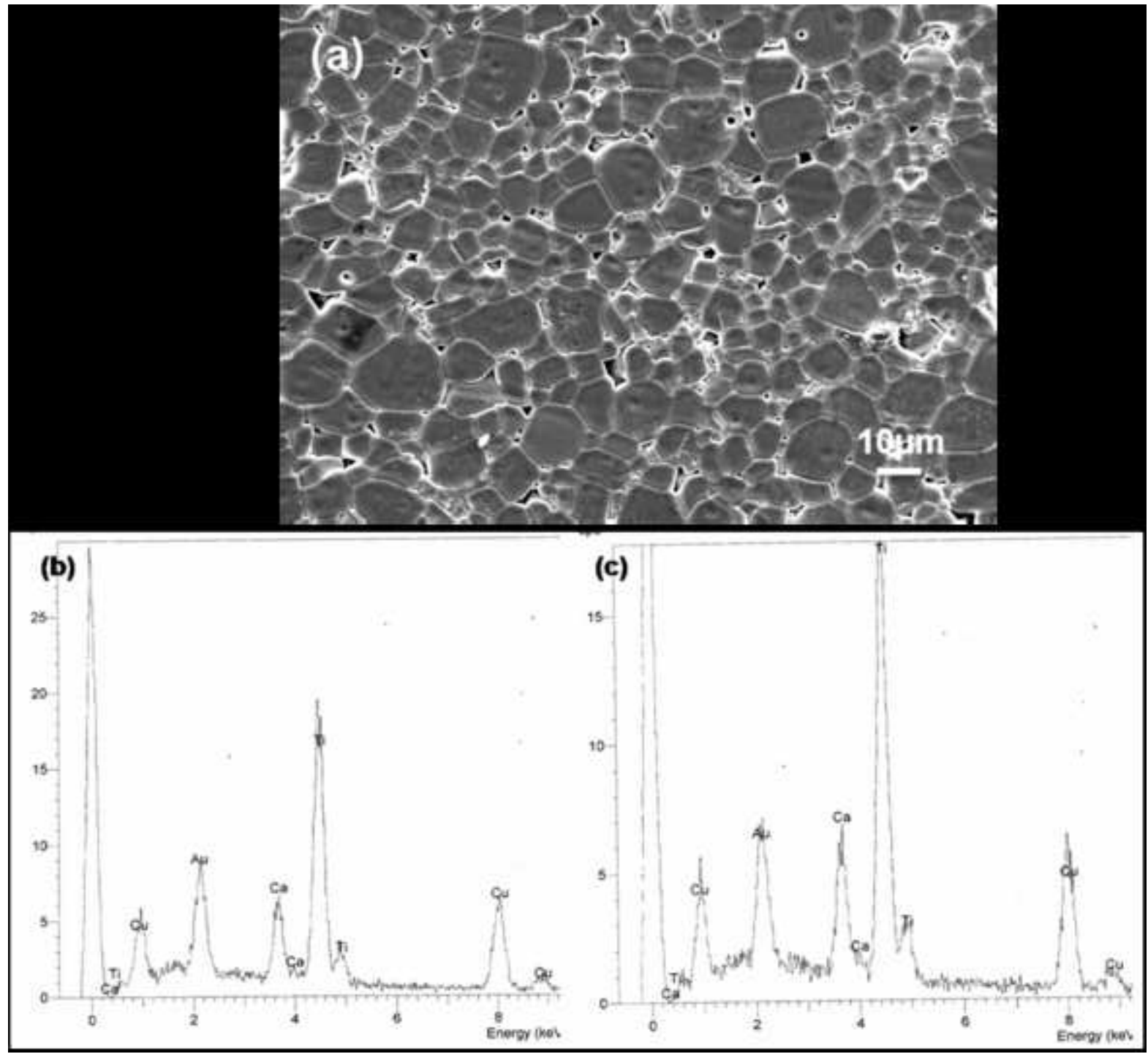

**Figure.5a**
**Click here to download high resolution image**

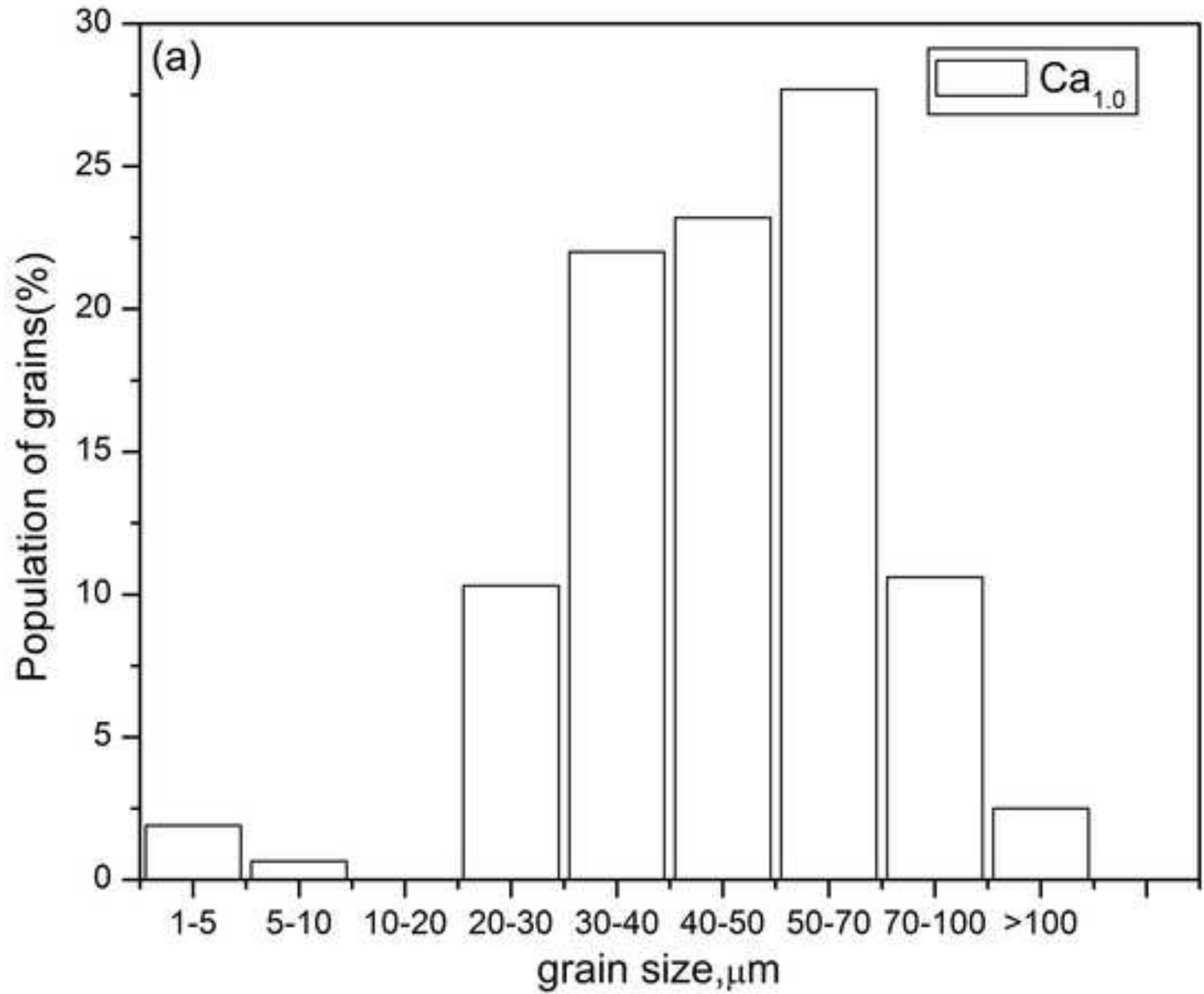

**Figure.5c**
**Click here to download high resolution image**

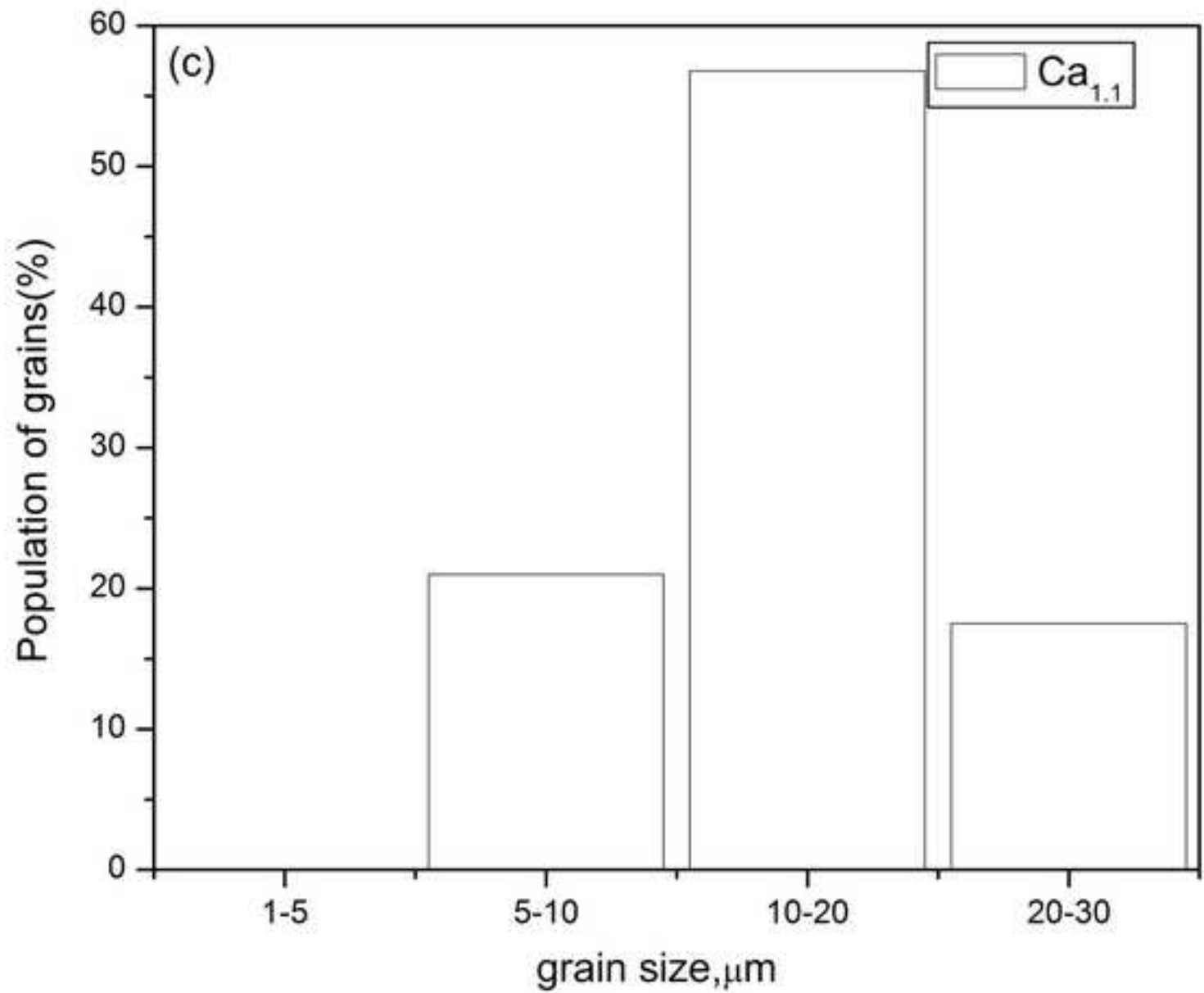

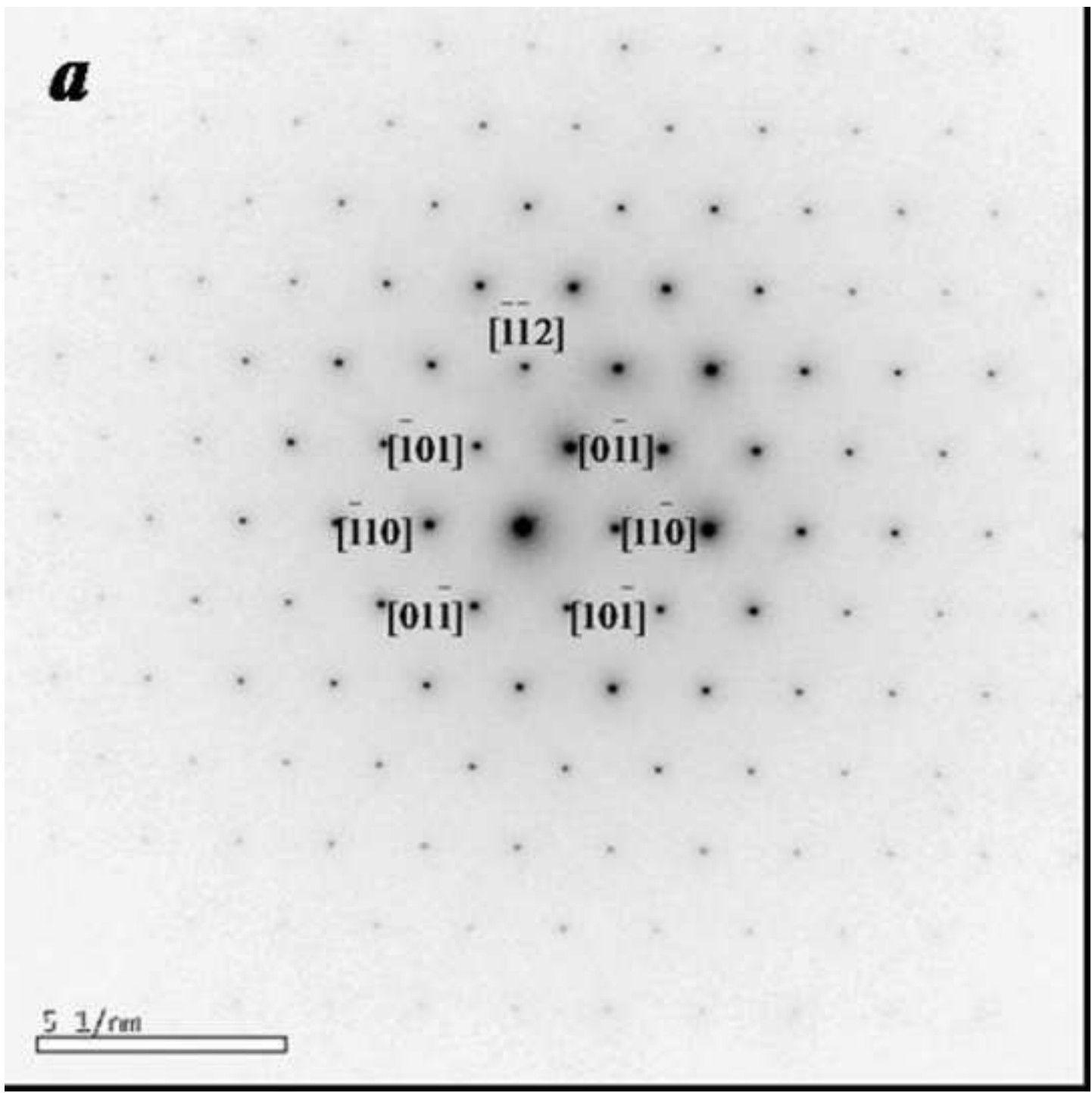

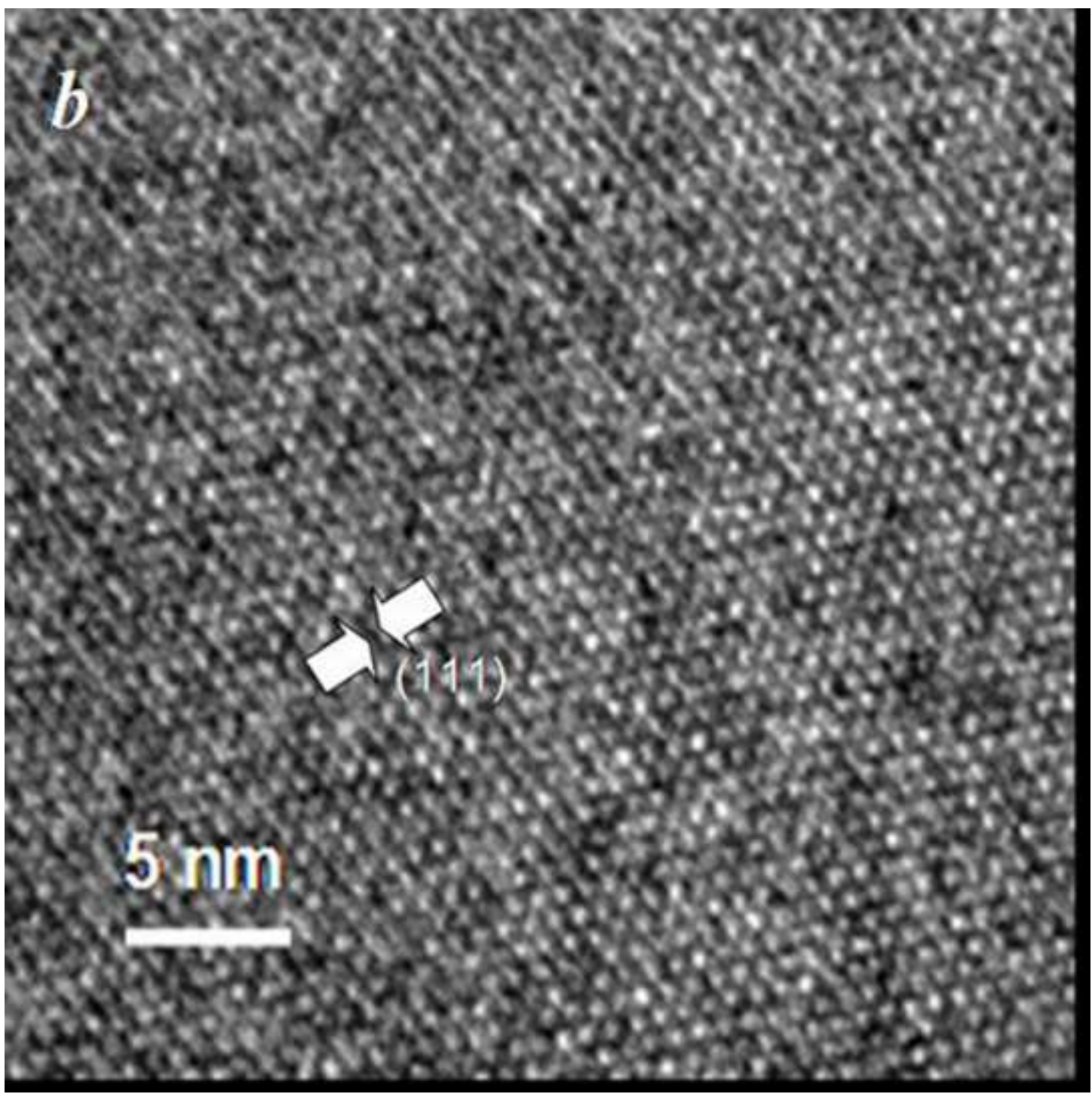

**Figure.7**
[Click here to download high resolution image](#)

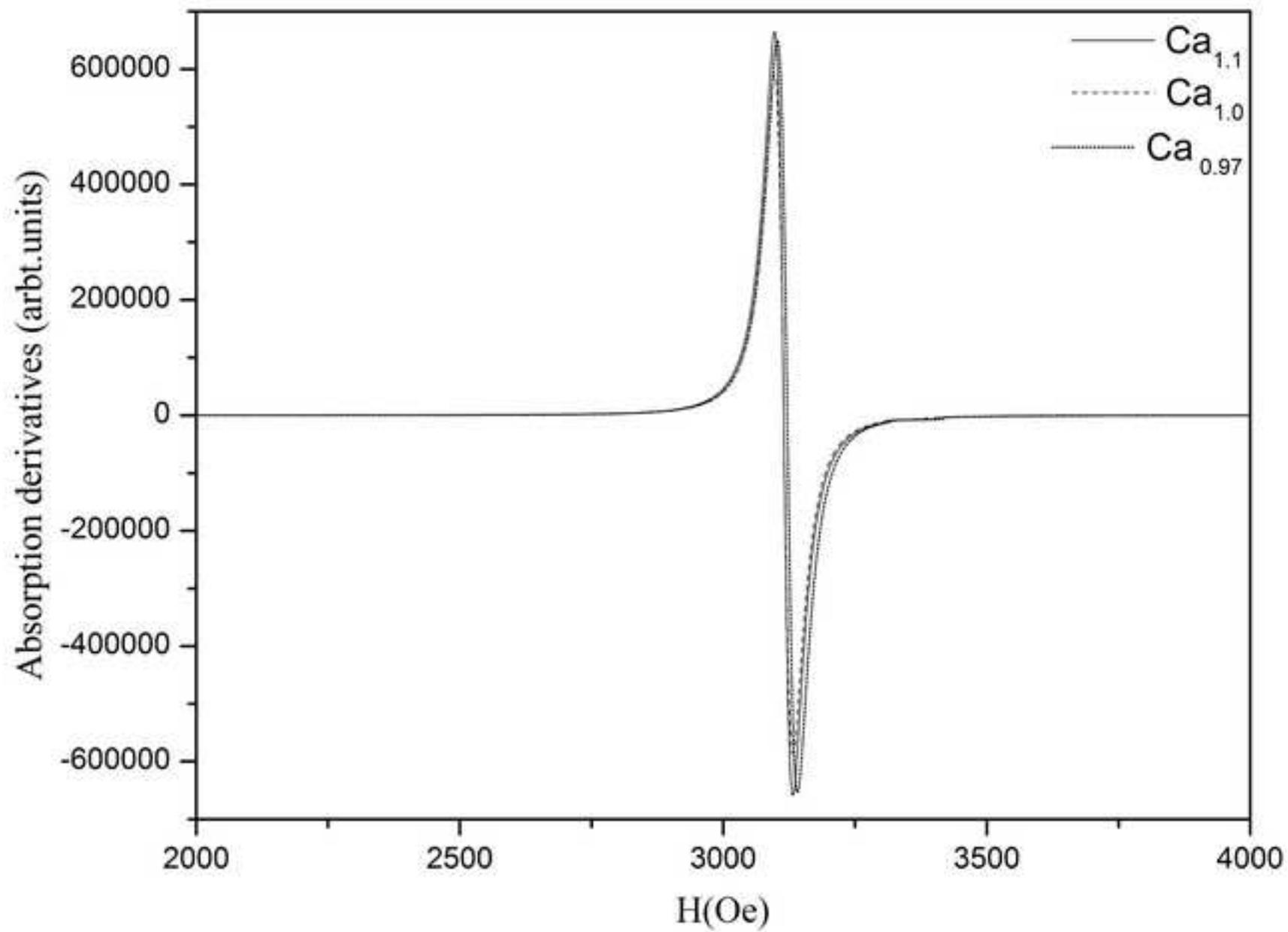

H(Oe)

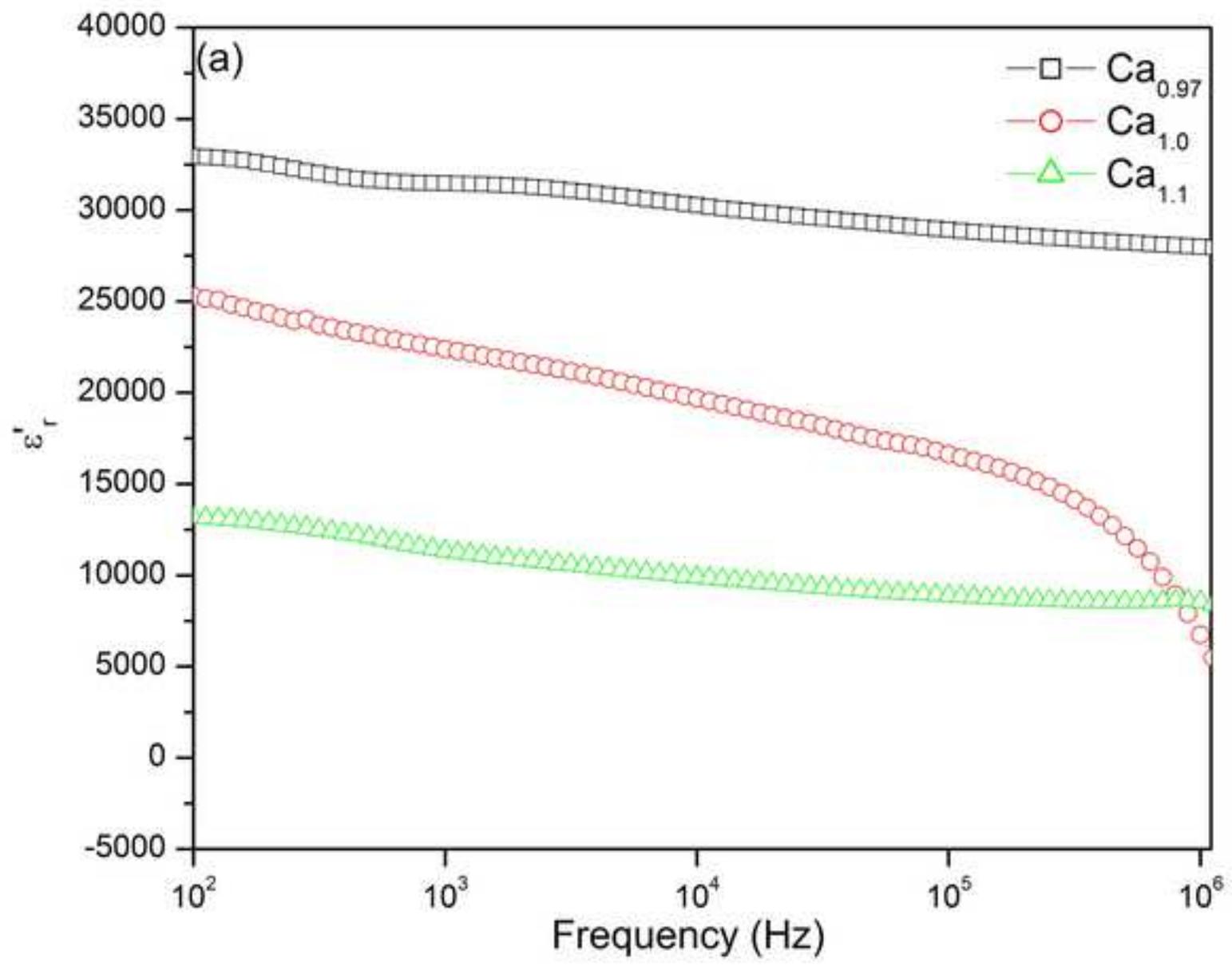

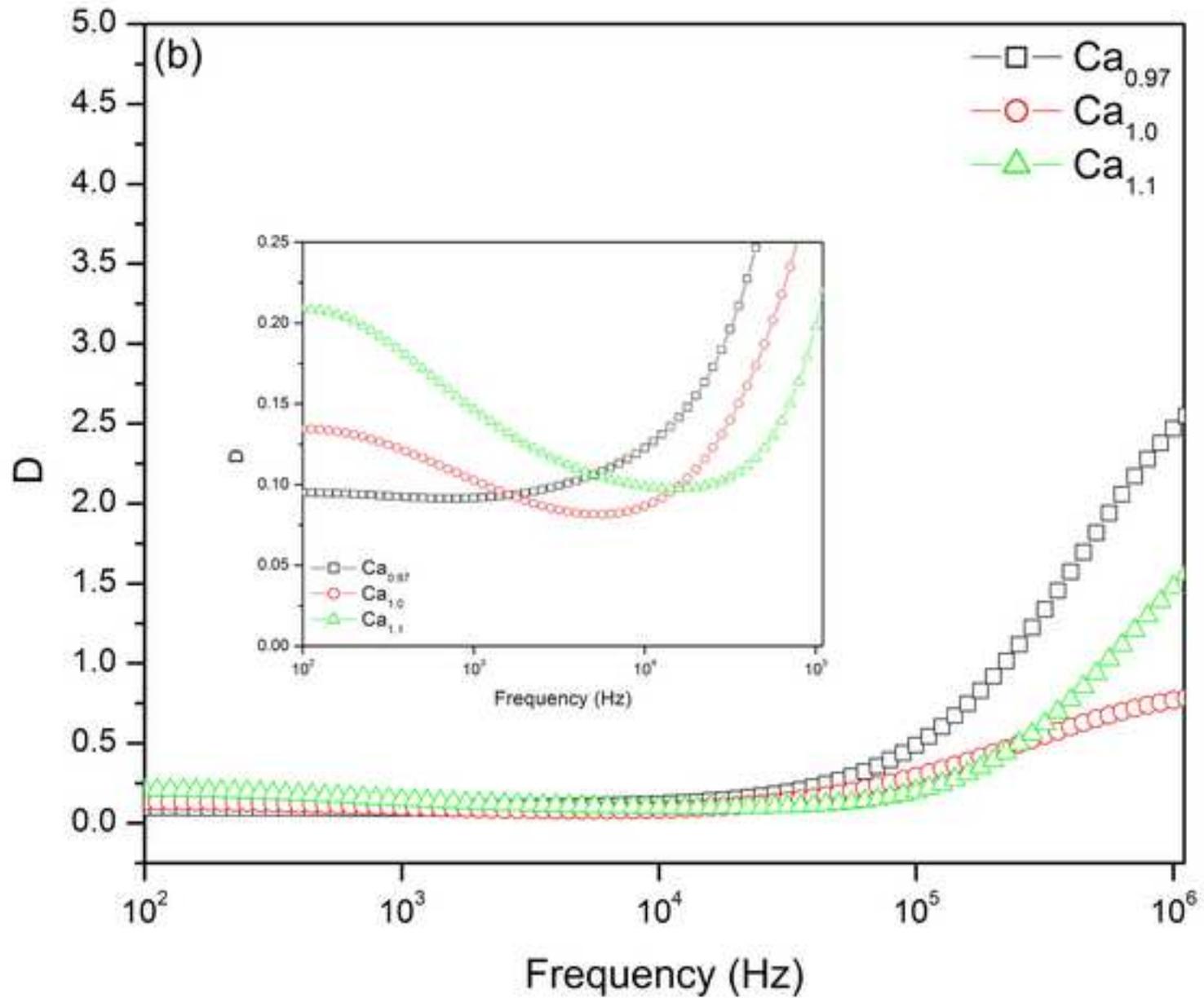

**Figure.9a**
**Click here to download high resolution image**

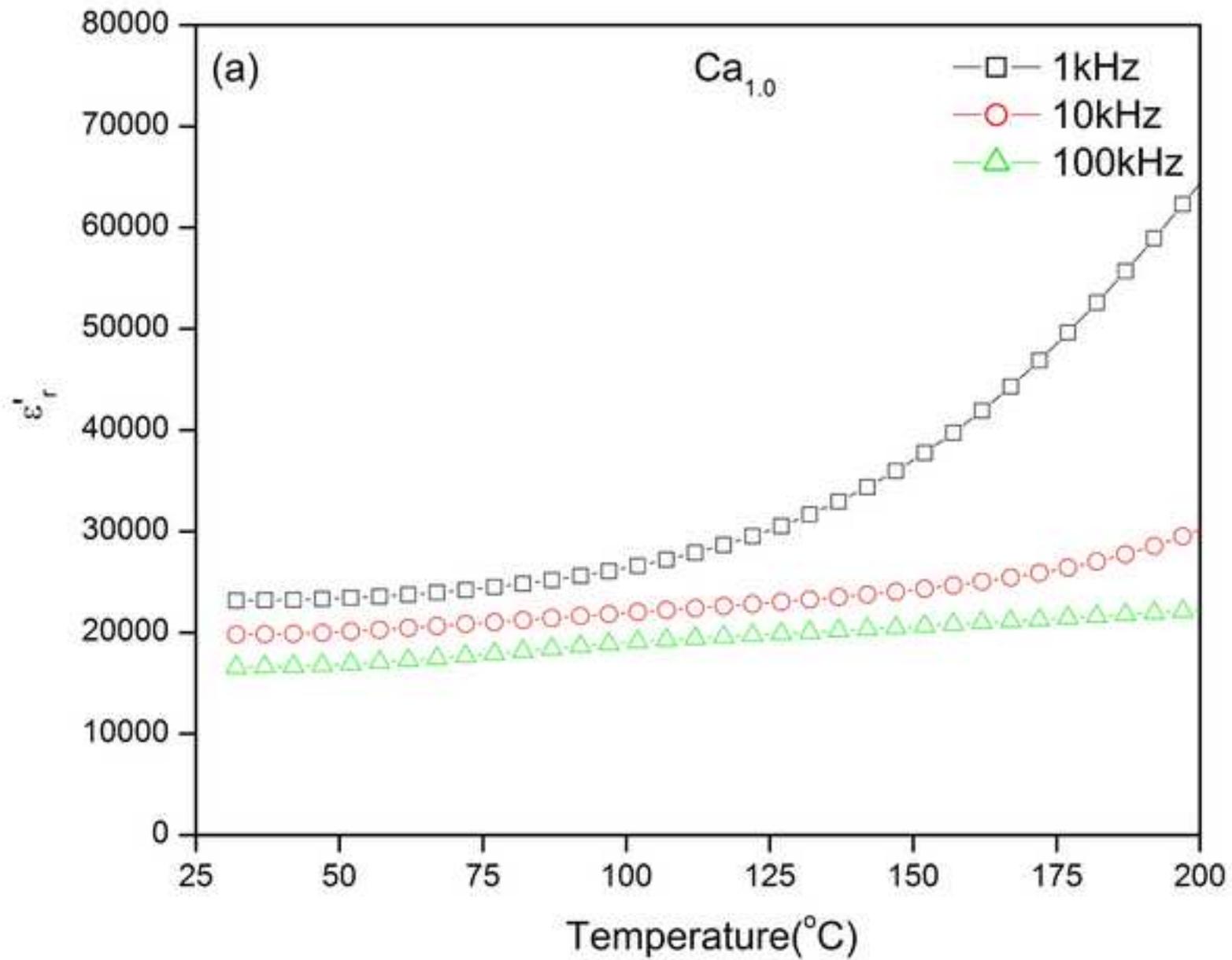

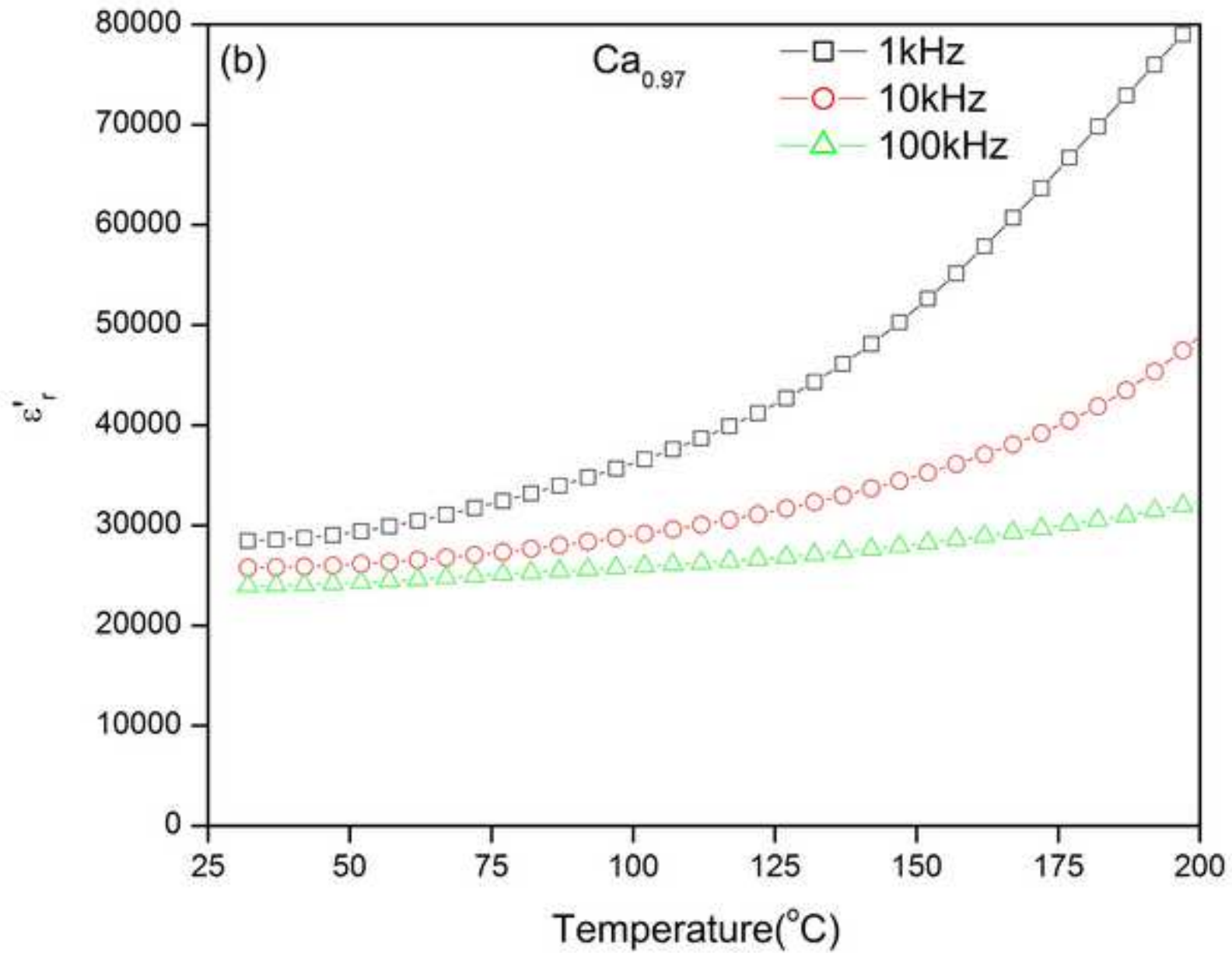

**Figure.9c**
**Click here to download high resolution image**

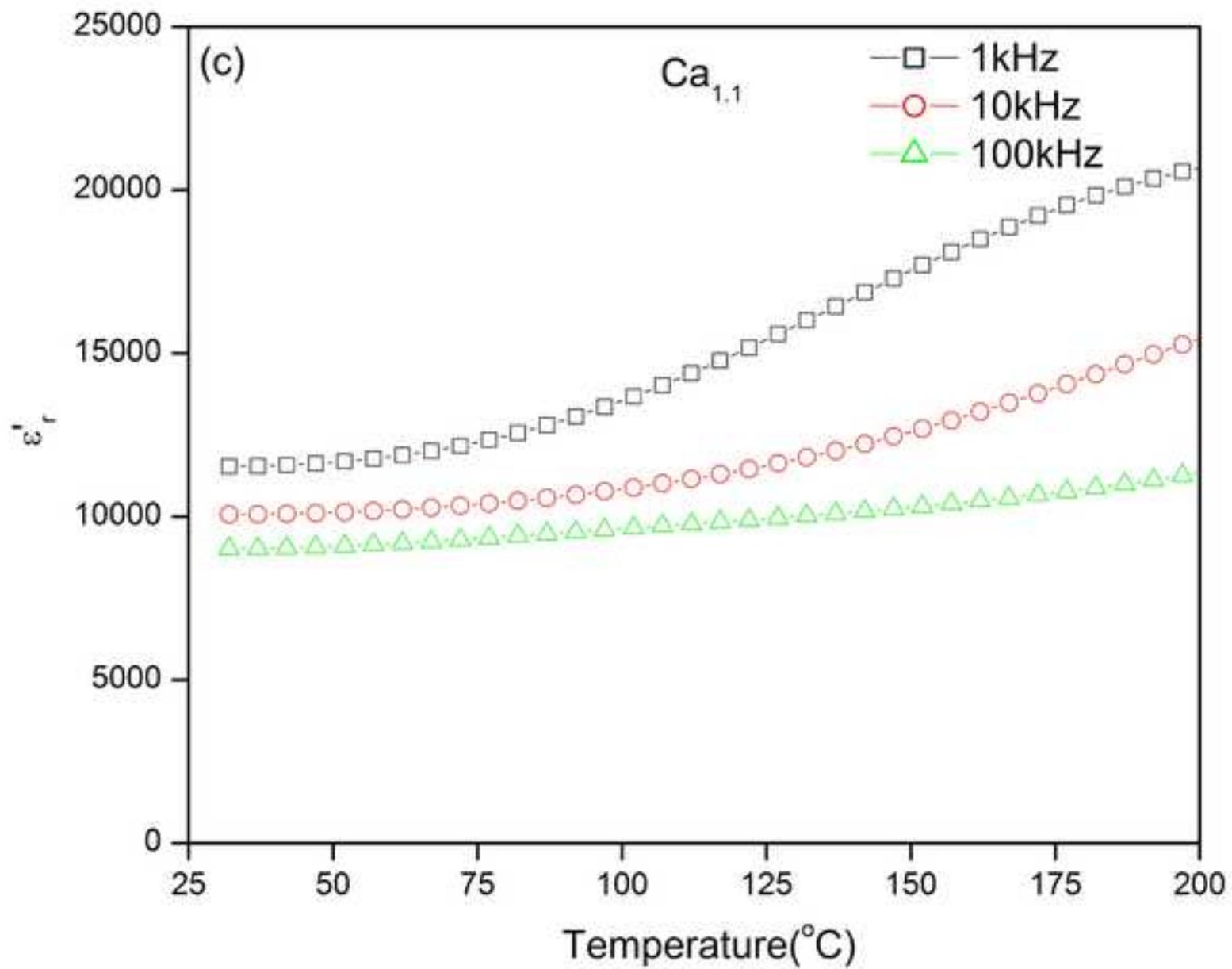

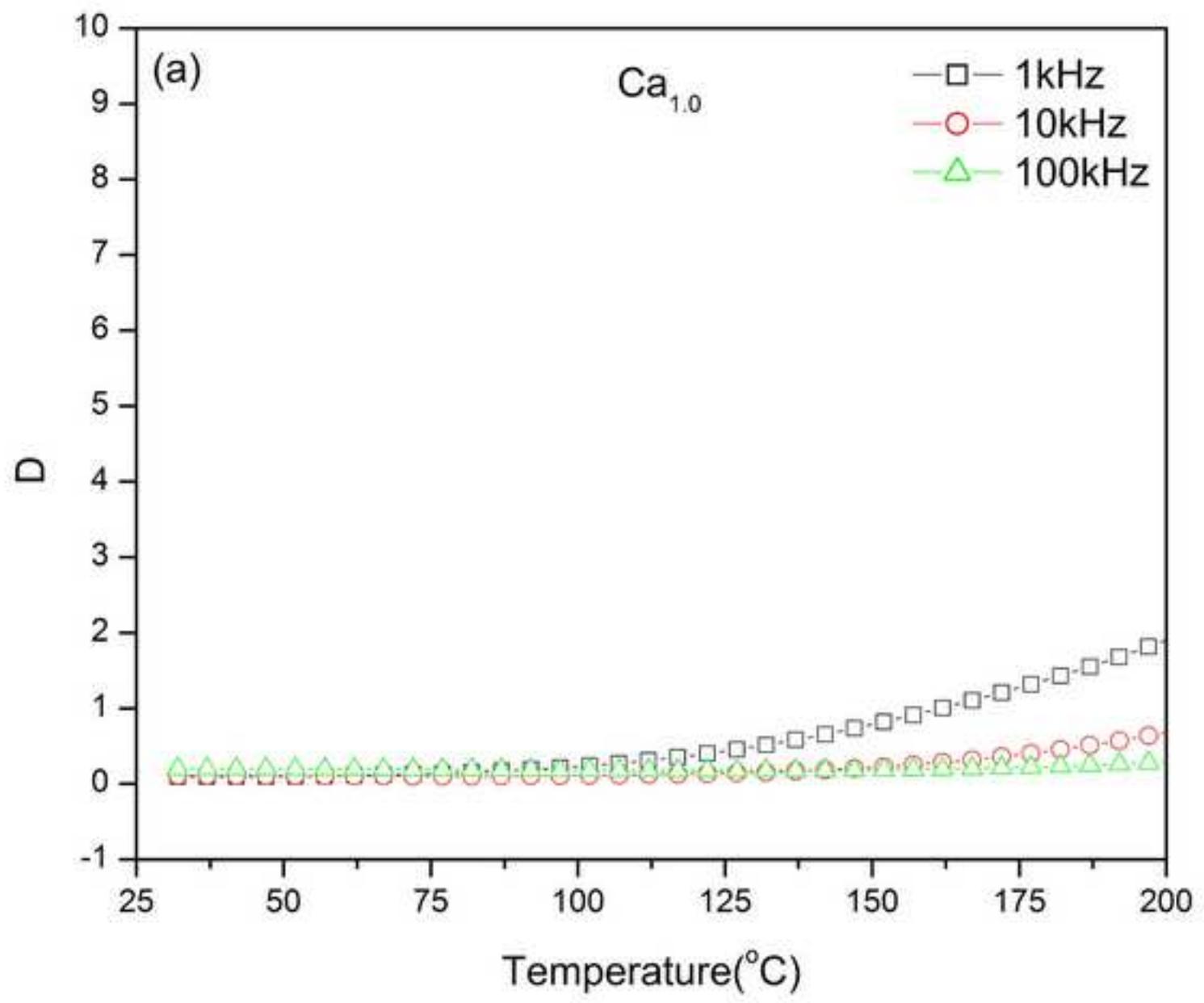

**Figure.10b**
**Click here to download high resolution image**

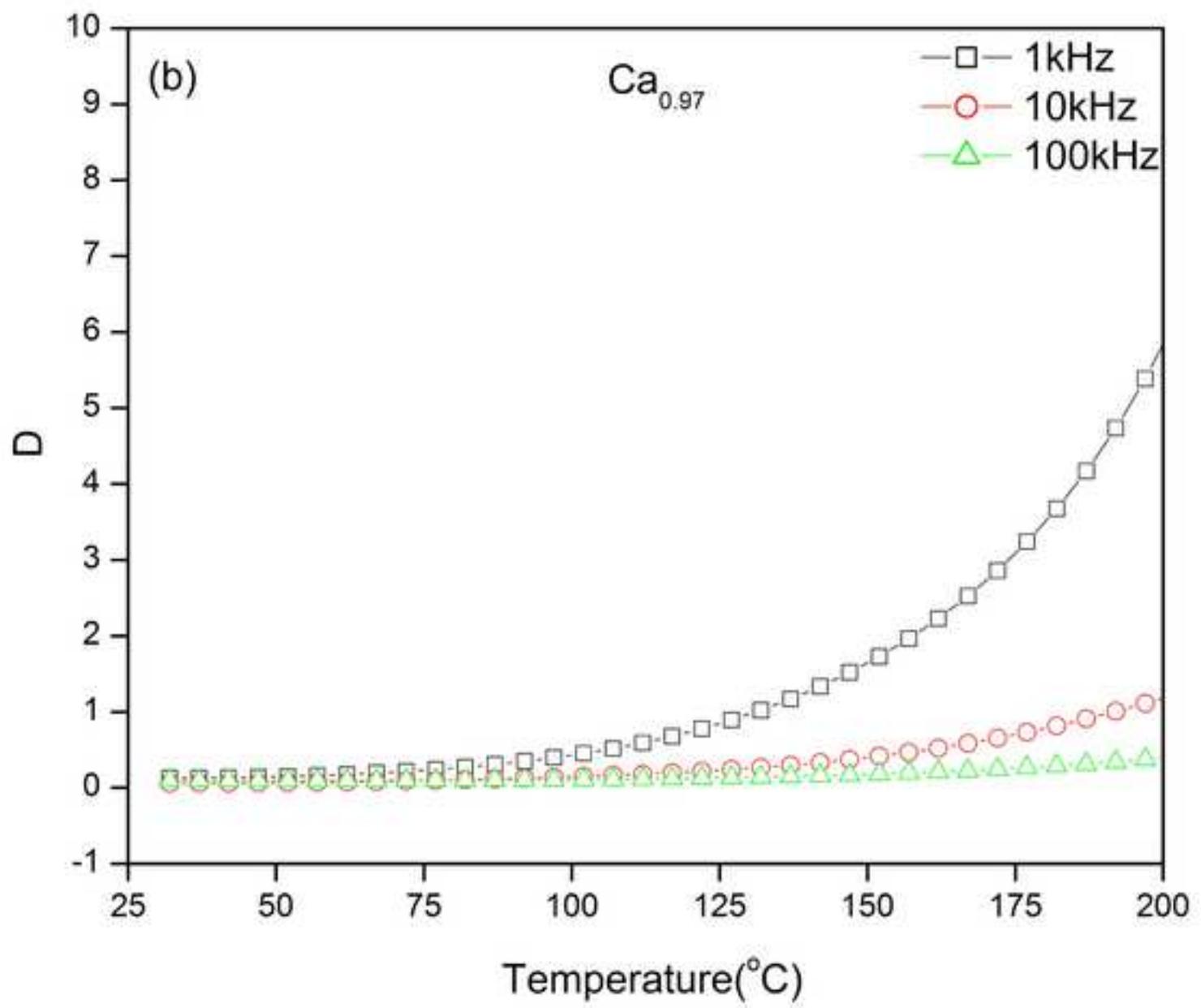

**Figure.10c**
**Click here to download high resolution image**

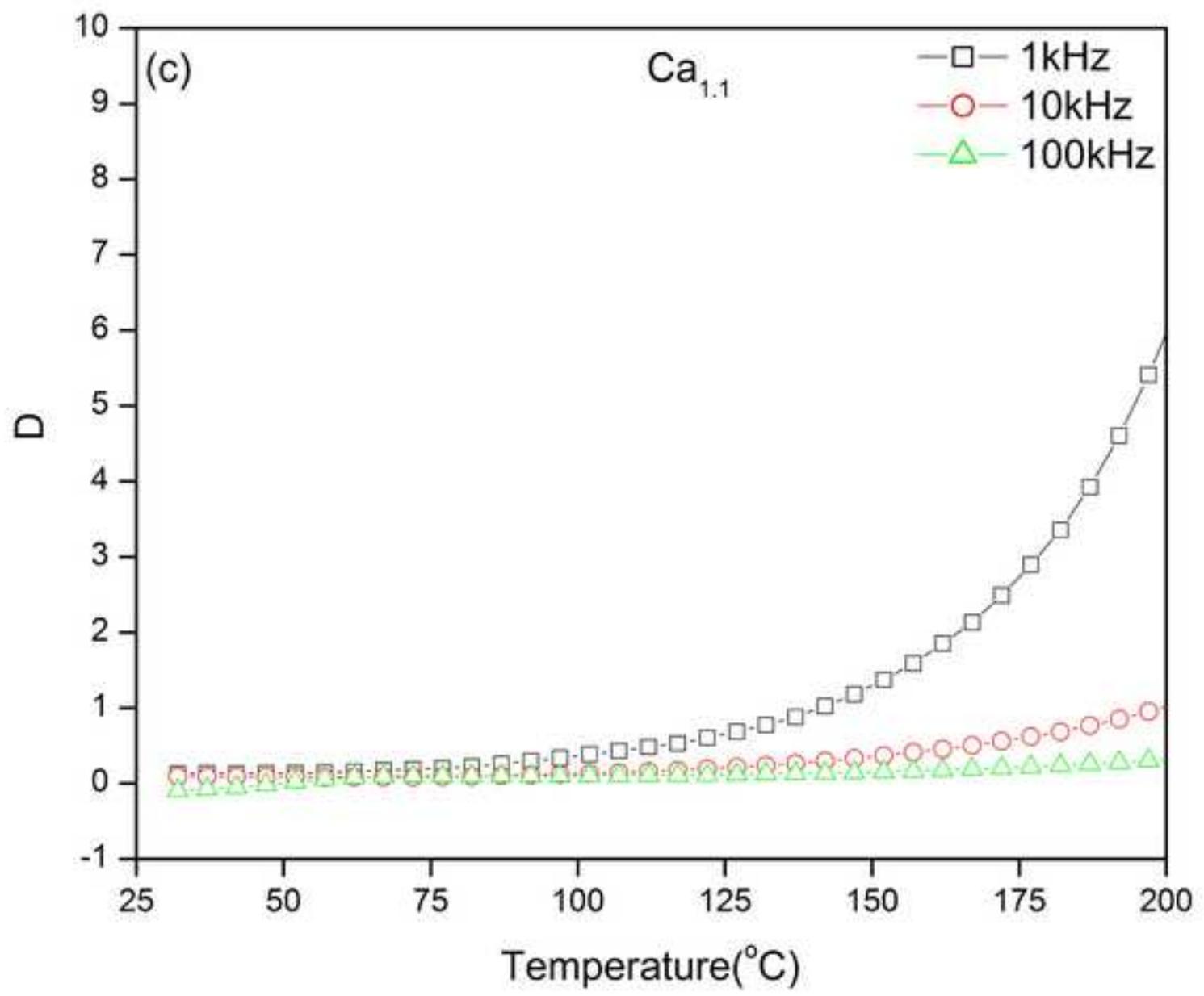

**Figure.5b**
**Click here to download high resolution image**

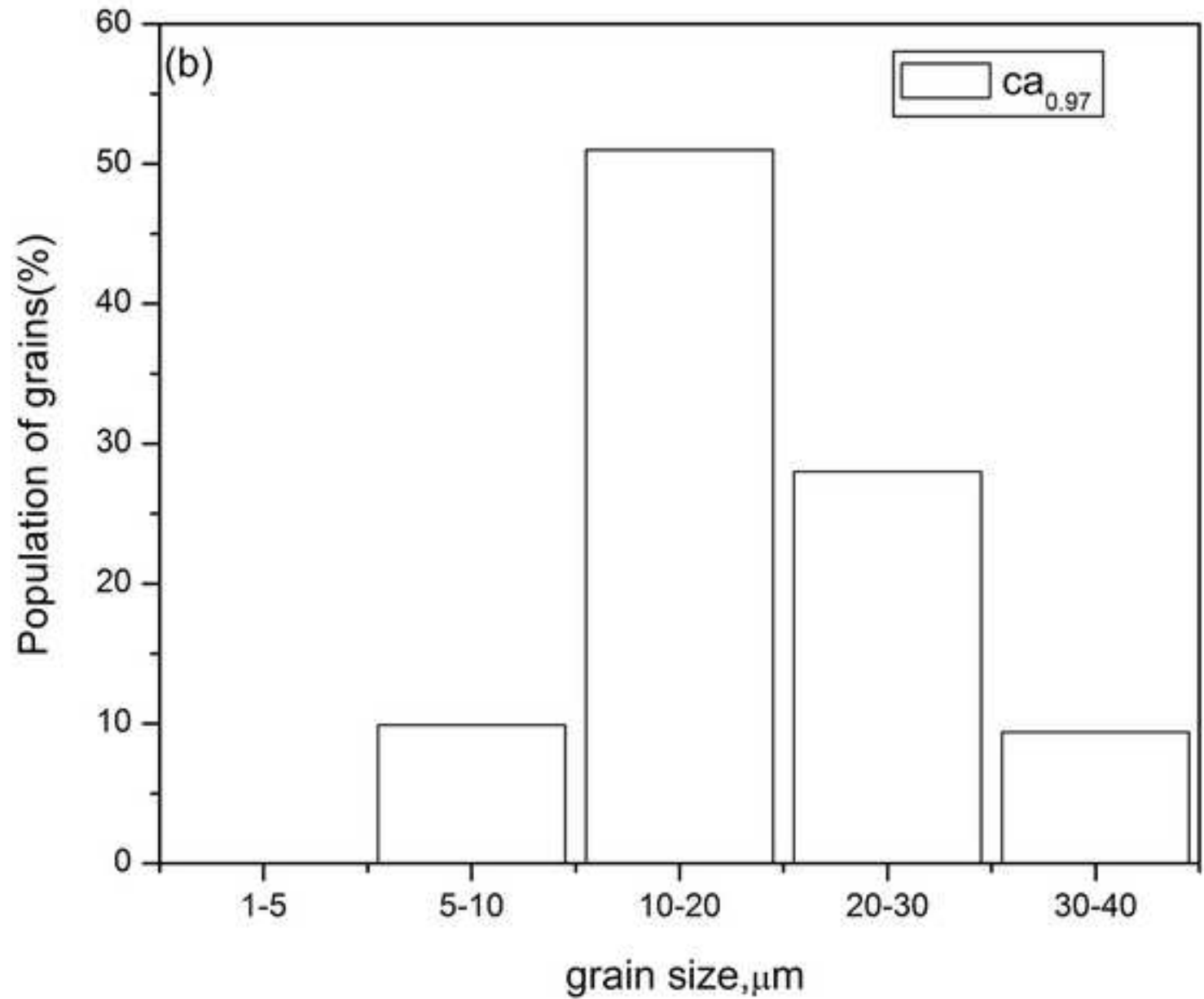

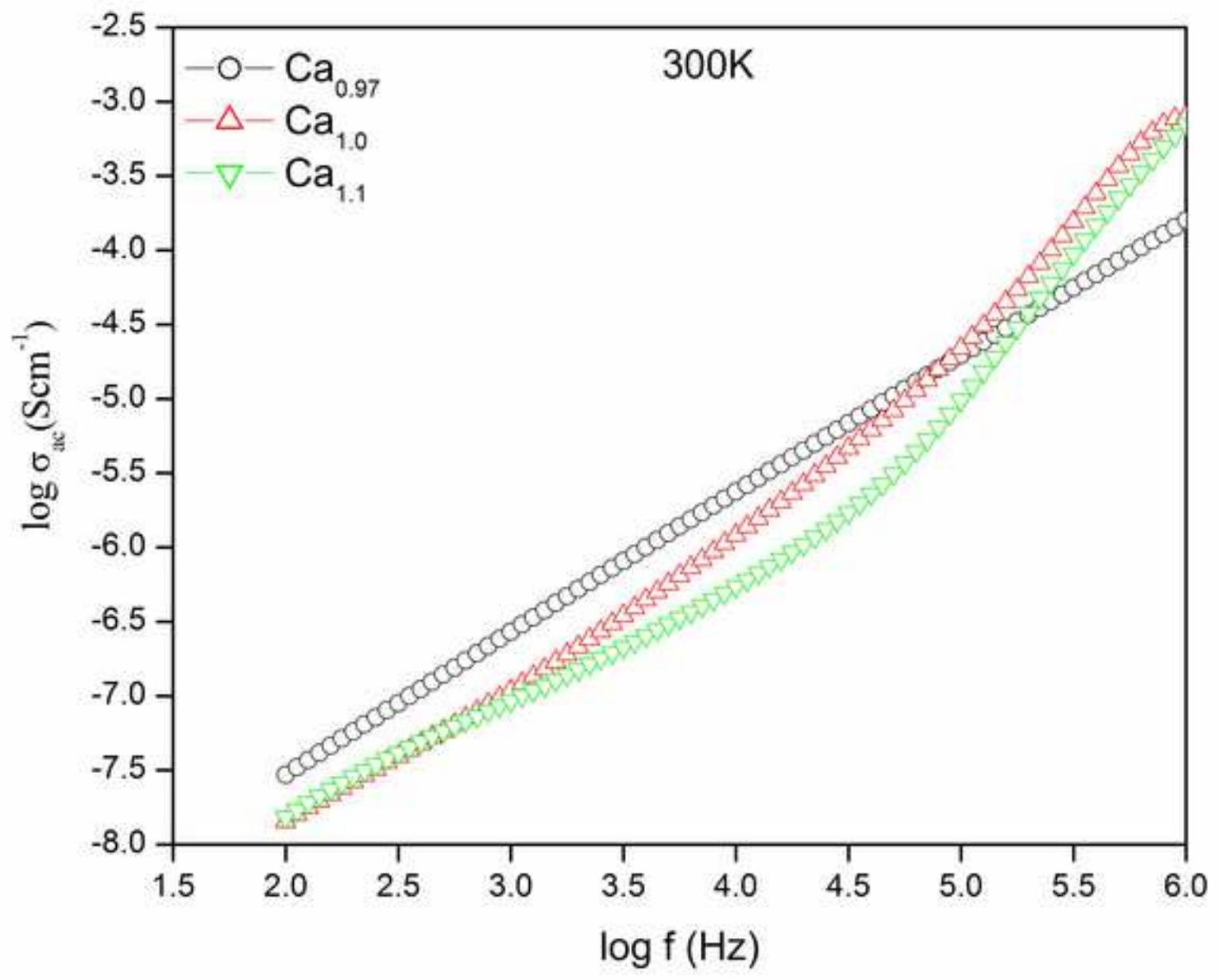